\documentclass[a4paper]{cas-sc}
\usepackage{subfigure}
\usepackage{comment}
\usepackage{xcolor}
\usepackage{graphicx}
\usepackage{float}
\usepackage{url}
\usepackage{lineno, hyperref}
\modulolinenumbers[1]

\begin{document}
\let\printorcid\relax
\shorttitle{}
\shortauthors{C. Lim et~al.}
\title [mode = title]{Development of a low-level radio frequency system for remote in situ superconducting radio frequency cavity measurements and calibrations in conduction-cooling systems}

\author[1]{\textcolor{black}{C. Lim}}
\author[1,3]{\textcolor{black}{K. Seol}}
\author[1]{\textcolor{black}{S. E. Park}}
\author[1,2]{\textcolor{black}{E.-S. Kim}}
\author[4]{\textcolor{black}{E. Kako}}
\author[2]{\textcolor{black}{S. Lee}}\cormark[1]
\author[1,2]{\textcolor{black}{B. R. Ko}}\cormark[2]

\affiliation[1]{organization={Department of Accelerator Science, Korea University Sejong Campus},
                addressline={2511 Sejong-ro}, 
                city={Sejong},
                citysep={},
                postcode={30019},
                country={Republic of Korea}}

\affiliation[2]{organization={Center for Accelerator Research, Korea University Sejong Campus},
                addressline={2511 Sejong-ro}, 
                city={Sejong},
                citysep={},
                postcode={30019},
                country={Republic of Korea}}

\affiliation[3]{organization={Institute for Rare Isotope Science, Institute for Basic Science},
                addressline={1 Gukjegwahak-ro},
                city={Yuseong-gu},
                postcode={Daejeon 34000}, 
                country={Republic of Korea}}

\affiliation[4]{organization={High Energy Accelerator Research Organization, KEK},
                addressline={1-1 Oho},
                city={Tsukuba},
                postcode={Ibaraki 305-0801}, 
                country={Japan}}

\cortext[cor1]{Corresponding author. Email : huvx44@korea.ac.kr}
\cortext[cor2]{Corresponding author. Email : brko@korea.ac.kr}

\begin{abstract}
  We developed a low-level radio frequency (LLRF) system for
  superconducting radio frequency (SRF) cavity measurements in a
  conduction-cooling system.
  A voltage-controlled oscillator (VCO) was used in a phase-locked
  loop (PLL) to track the frequencies of SRF cavity modes.  
  {\tt{Python}}-based software developed in-house controlled and
  monitored the system.
  By introducing an additional circulator and associated components,
  we developed an LLRF system enabling fully remote in situ SRF cavity
  measurements for the first time.
  This design also enables remote in situ calibrations.
  Hence, our LLRF system improves safety through fully remote
  operation and enhances measurement precision through in situ
  calibrations, with safety always taking priority. These calibrations
  are important for measurements in conduction-cooling systems because
  temperature stability and uniformity in these systems are poorer
  than those in convection-cooling systems using costly liquid helium
  and can vary over time and with environmental conditions.
\end{abstract}

\begin{keywords}
LLRF \sep VCO-PLL \sep SRF cavity \sep remote in situ measurements \sep remote in situ calibrations \sep conduction cooling
\end{keywords}

\maketitle

\section{Introduction}\label{sec:intro}
Superconducting radio frequency (SRF) cavities have been widely and
successfully employed in modern accelerators.
Specifically, high-purity Nb-based SRF cavities have achieved a
state-of-the-art accelerating gradient ($E_{\rm acc}$) at a physical
temperature of approximately 2~K~\cite{TESLA} and have thus been
employed in current cutting-edge high-energy
accelerators~\cite{LHC1,LHC2,SKEKB1,SKEKB2} for fundamental
science~\cite{ATLAS,CMS,Belle2}; they will also be used in future
accelerators~\cite{ILC,CEPC,FCC}. However, operation at 2~K requires
both large-scale cryogenic infrastructure and many cryogenics experts,
making operation outside national laboratories very difficult. Beyond
fundamental research at world-leading accelerator facilities, particle
accelerators have many environmental, medical, materials science, and
other applications~\cite{ACCAPP1,ACCAPP2,ACCAPP3}. Such applications
require megaelectron-volt-scale beam energy and thus can be realized
with an SRF cavity only approximately 1~m long at an appropriate
operating temperature. Because of their small size, SRF cavities in
particle accelerators for applied research can be conduction-cooled
using a cryogen-free cryocooler without large-scale cryogenic
infrastructure requiring costly liquid helium (LHe). Furthermore,
Nb$_3$Sn-based SRF cavities can, in principle, achieve performance
comparable to that of Nb SRF cavities operated at approximately 2~K,
even at approximately 4~K~\cite{Nb3Sn1,Nb3Sn2}. These Nb$_3$Sn-based
cavities can therefore achieve optimal performance even within typical
cryocooler specifications; consequently, research and development of
accelerators using conduction-cooled Nb$_3$Sn-based SRF cavities is
expanding worldwide, garnering considerable interest across many
fields of applied
science~\cite{CCFermilab1,CCFermilab2,CCJefferson,CCPKU,CCIMP,MICROPHONICS}.

Unlike a convection-cooling system, a typical conduction-cooling
system operates without requiring immersion in LHe and thus exhibits
temperature gradients and greater susceptibility to ambient
temperature fluctuations. Hence, in situ measurements and calibrations
are desirable for precise SRF cavity measurements in
conduction-cooling systems and must be performed remotely to avoid
exposure to hazardous radiation.

To address these requirements, we developed a low-level radio
frequency (LLRF) system for SRF cavity measurements in a
conduction-cooling system. The system uses a voltage-controlled
oscillator (VCO) in a phase-locked loop (PLL)~\cite{VCOPLL} to track
the frequencies of the accelerating gradient of SRF cavities. It also
enables remote in situ measurements and calibrations using a
radio frequency (RF) circulator in a low-power region, an approach inspired
by the use of a circulator in a high-power region. \verb|Python|-based
software developed in-house controls the system and manages data
acquisition (DAQ).
\section{LLRF system for SRF cavity measurements}\label{sec:llrf}
\begin{figure}[h]
  \centering
  \includegraphics[width=0.75\textwidth]{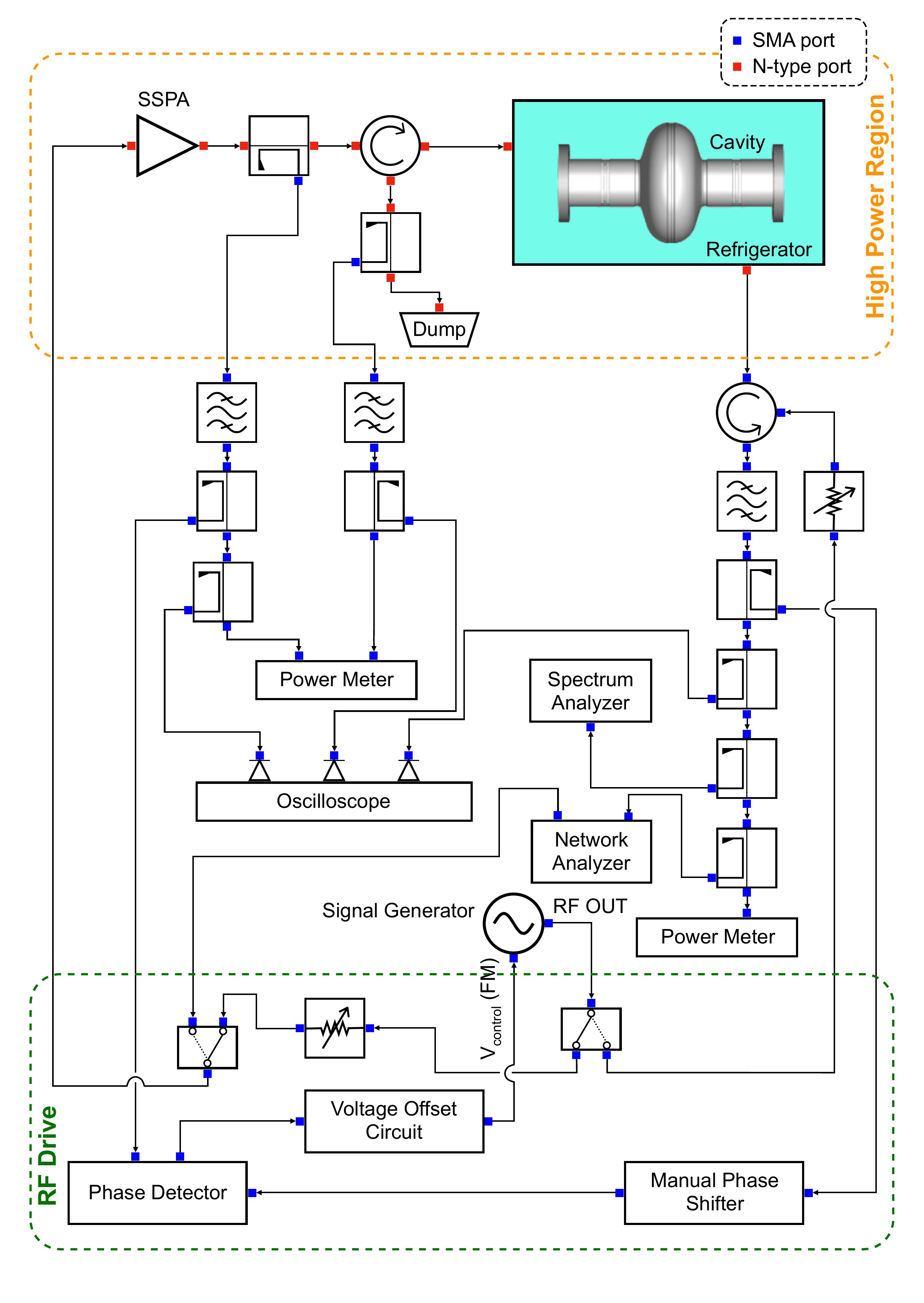}
  \caption{Block diagram of the LLRF system. By incorporating the RF
    switches and variable attenuator, along with an additional
    circulator immediately after the cavity output port, we developed
    an LLRF system enabling fully remote in situ SRF cavity
    measurements for the first time.}
  \label{FIG:WHOLELLRF}
\end{figure}
The LLRF system was developed primarily to verify the RF
characteristics of SRF cavities, such as the unloaded quality factor
of the cavity mode ($Q_0$) as a function of the accelerating gradient
($E_{\rm acc}$).
\begin{figure}[h]
  \centering
  \includegraphics[width=0.72\textwidth]{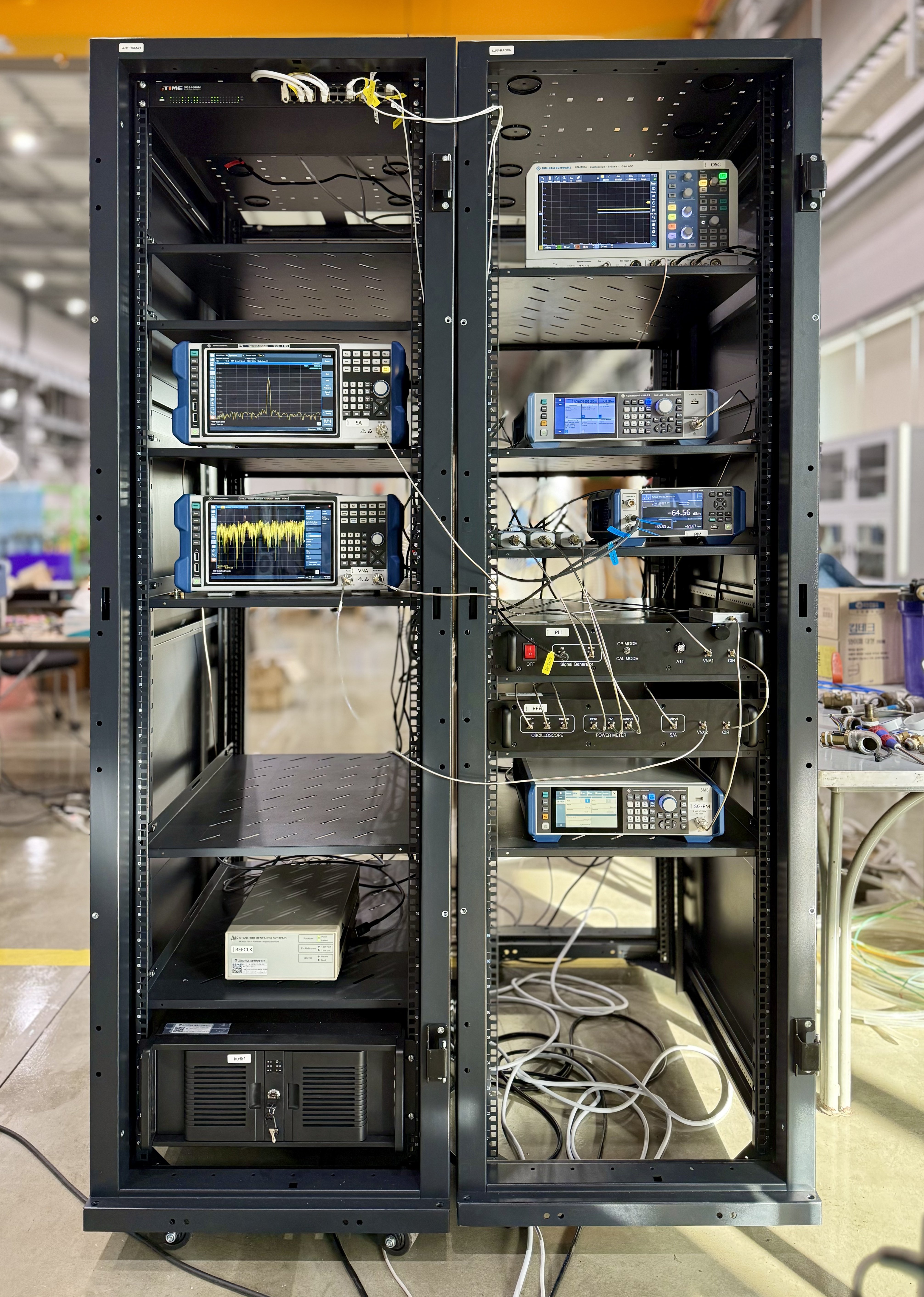}
  \caption{Part of the constructed LLRF system corresponding to the
    block diagram in Figure~\ref{FIG:WHOLELLRF}. From top to bottom,
    the left rack contains a network switch, a spectrum analyzer, a
    network analyzer, a frequency standard, and the control computer.
    From top to bottom, the right rack contains an oscilloscope, a
    signal generator to simulate the cavity mode frequency, power
    meters, RF boxes (depicted in Figure~\ref{FIG:LLRF_BOTH}), and
    another signal generator operating in frequency-modulation
    mode. The setup is currently configured for the PLL test discussed
    in Section~\ref{sec:pll}; the spectrum analyzer displays a sharp
    peak at the frequency locked by our LLRF system, and the
    oscilloscope displays an output resulting from the voltage
    offset.}  
  \label{FIG:LLRF_RACK}
\end{figure}
Figure~\ref{FIG:WHOLELLRF} presents the block diagram of our LLRF
system, and Figure~\ref{FIG:LLRF_RACK} depicts its physical
implementation, excluding the high-power region.

Because SRF cavities have extremely high $Q_0$ values and thus very
narrow resonance bandwidths, typically in the sub-hertz range,
applying an RF signal precisely at the cavity's resonant frequency is
particularly important for exciting its resonant mode. $Q_0$ can be
determined from the relation
\begin{eqnarray}
  Q_0    &=& (1+\beta_{\rm in}+\beta_{\rm out})Q_L, 
  \label{EQ:Q0}
\end{eqnarray}
where $Q_L$ is the loaded quality factor. $\beta_i$  is the coupling
coefficient for the relevant cavity port $i$ (input (in) or output
(out)) and is defined as
\begin{eqnarray}
  \beta_i\equiv\frac{P_i}{P_0}=\frac{Q_0}{Q_i},
  \label{EQ:BETA0}
\end{eqnarray}
where $P_0$ is the power dissipated on the cavity wall by the relevant
cavity mode, and $P_i$ and $Q_i$ are the power loss and quality factor,
respectively, of the relevant cavity port $i$. The second relation in
Equation~(\ref{EQ:BETA0}) follows from the definition of the quality
factor, $Q_k\equiv\omega_0 U/P_k$, where $Q_k$ denotes the quality
factor of system component $k$ (the cavity or a port), and $\omega_0$
and $U$ are the resonant frequency and total energy of the system,
respectively.

According to refs.~\cite{BETA1, BETA2} and Equation~(\ref{EQ:BETA0}),
Equation~(\ref{EQ:Q0}) assumes that $\beta_{\rm in}$ and
$\beta_{\rm out}$ are independent of each other, as if each were
measured in the absence of the other, i.e., in a single-port
system. However, when both ports are present, i.e., in a two-port
system, the coupling coefficients used to estimate $Q_0$ from
Equation~(\ref{EQ:Q0}) are no longer independent and are given by
\begin{eqnarray}
  \beta_{\rm in}  &=&\frac{\beta^*_{\rm in}(1+\beta^*_{\rm out})}{1-\beta^*_{\rm in}\beta^*_{\rm out}} \nonumber \\
  \beta_{\rm out} &=&\frac{\beta^*_{\rm out}(1+\beta^*_{\rm in})}{1-\beta^*_{\rm in}\beta^*_{\rm out}},
  \label{EQ:BETA-2PORT}
\end{eqnarray}
where $\beta^*_{\rm in}$ and $\beta^*_{\rm out}$ are the coupling
coefficients of the cavity input and output ports, respectively, as
measured in a two-port system~\cite{BETA1, BETA2}.
Equation~(\ref{EQ:BETA-2PORT}) is valid when
$\beta^*_{\rm in}\beta^*_{\rm out}<1$, a condition satisfied by the
conventional coupling configuration used in accelerators:
$\beta^*_{\rm in}\approx1$ for maximum power delivery to the cavity
and $\beta^*_{\rm out}\ll1$ for monitoring and testing without
extracting excessive energy stored in the cavity mode. The LLRF system
should be able to measure $Q_L$, $\beta^*_{\rm in}$,
and $\beta^*_{\rm out}$ accurately to determine $Q_0$, which requires
precisely matching the RF signal's frequency to the cavity's resonant
frequencies. Because the SRF cavity's resonant frequency can vary
slightly over time due to many factors that depend on experimental and
environmental conditions, identifying its changing resonant
frequencies can be challenging but is crucial for accurately
determining its $Q_0$. This RF drive is described in
Section~\ref{sec:pll}.

$E_{\rm acc}$ is among the most important parameters of accelerators
and can be determined from the relation
\begin{eqnarray}
    E_{\rm acc}&=&\frac{1}{l_c}\sqrt{\frac{R_{\rm sh}}{Q_0}}\sqrt{P_0 Q_0} \nonumber \\
               &=&\frac{1}{l_c}\sqrt{\frac{R_{\rm sh}}{Q_0}}\sqrt{P_{\rm in} Q_{\rm in}}=\frac{1}{l_c}\sqrt{\frac{R_{\rm sh}}{Q_0}}\sqrt{P_{\rm out} Q_{\rm out}},
    \label{EQ:ECC}
\end{eqnarray}
where $l_c$ is the cavity length and $R_{\rm sh}$ is the shunt
resistance. The second line of Equation(\ref{EQ:ECC}) follows from
Equation~(\ref{EQ:BETA0}), which yields $P_0 Q_0=P_i Q_i$. Measuring
$P_0$ is not straightforward, and measuring $P_{\rm in}$ is more
challenging overall than measuring $P_{\rm out}$; consequently,
$P_{\rm out}$ measurements are generally used to obtain the
accelerating gradient $E_{\rm acc}$. Furthermore, our LLRF system
enabled remote in situ calibrations for more precise $P_{\rm out}$
measurements and thus more precise $E_{\rm acc}$ measurements;
achieving such remote in situ measurements and calibrations via the
cavity input port are also difficult. These remote in situ
measurements and calibrations are discussed in
Sections~\ref{sec:power} and \ref{sec:calib}, respectively.
\subsection{RF drive}\label{sec:pll}
\begin{figure}[h]
  \centering
  \includegraphics[width=0.85\textwidth]{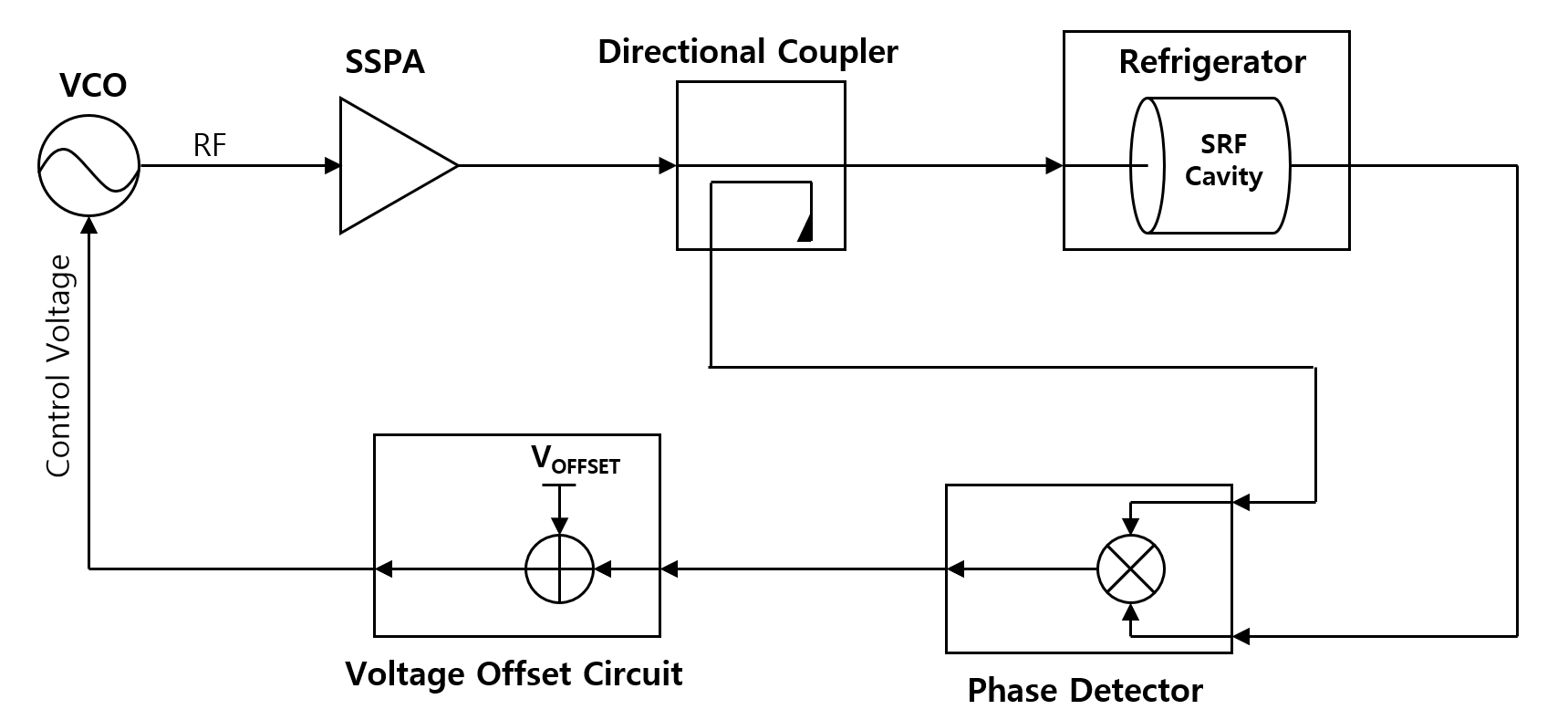}
  \caption{Block diagram corresponding to the PLL.}
  \label{FIG:PLL}
\end{figure}
Applying a matched RF signal is critical for both verifying the RF
characteristics of SRF cavities and operating these cavities in an
actual accelerator system.
\begin{figure*}[h]
  \centering
  \subfigure{\includegraphics[width=0.80\textwidth]{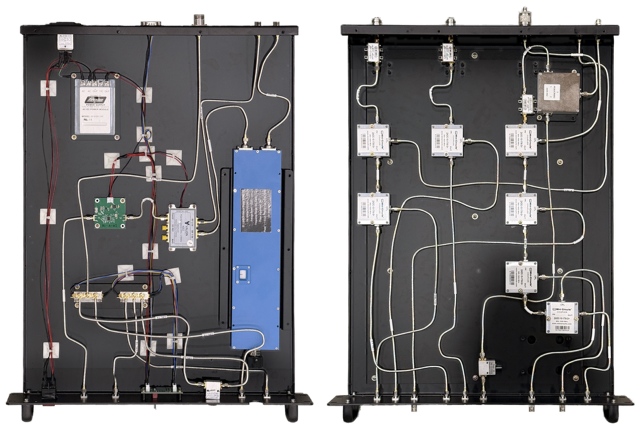}}
  \caption{Parts of the constructed LLRF system. The left panel
    corresponds to the RF drive in Figure~\ref{FIG:WHOLELLRF}, whereas
    the right panel corresponds to the part of the LLRF system in
    Figure~\ref{FIG:WHOLELLRF} that excludes the RF drive, high-power
    region, and RF equipment (spectrum analyzer, network analyzer,
    oscilloscope, power meters, and signal generator).
    The low-power circulator at the upper right of the right panel and
    the two RF switches just below the middle-left area of the left
    panel were introduced for the first time in this study for the
    remote in situ measurements and calibrations described in
    Sections~\ref{sec:power} and \ref{sec:calib}, respectively. They
    are now mounted in the right-hand rack depicted in
    Figure~\ref{FIG:LLRF_RACK}.}
  \label{FIG:LLRF_BOTH}
\end{figure*}
To address this, many existing accelerator facilities employ various
automatic tuning schemes~\cite{PLL, SEL, IRIS} to match the input
frequency to the resonant frequency of the relevant SRF cavity
mode. Our LLRF system was designed to track the frequency of the SRF
cavity mode using a PLL approach similar to those described in
refs.~\cite{PLL, IRIS}. The PLL mode offers high stability and precise
control of the power and phase of the output from the solid-state
power amplifier (SSPA), making it well-suited for accurate
measurements of conduction-cooled SRF cavities.

Figure~\ref{FIG:PLL} presents the configuration of the designed PLL
block, and the left panel of Figure~\ref{FIG:LLRF_BOTH} depicts part
of the constructed PLL. The RF signal generated by the VCO is
amplified by the SSPA and applied to the cavity. A mismatch between
the frequencies of the signal coupled to the cavity input and the
signal from the resonant mode inside the cavity results in a phase
difference. The AD8302-based phase detector converts this phase
difference into a direct-current voltage of up to 1.8~V, depending on
the operating frequency~\cite{AD8302}. The phase detector output then
passes through a voltage-offset circuit to match the control voltage
to the input voltage range of the VCO, which accepts
alternating-current voltages. The output of the voltage-offset circuit
ranges from $-0.9$ to $+0.9$~V and is fed back to the VCO as a control
voltage; this range corresponds to the phase detector's output range
of 0$\--$1.8~V shifted by $-0.9$~V. Consequently, the RF signal
frequency is adjusted to minimize the phase difference by the feedback
loop, thereby automatically aligning the two frequencies.

The network analyzer, together with the relevant RF switch depicted in
Figure~\ref{FIG:WHOLELLRF}, was used both to provide the PLL with an
initial seed frequency for the cavity's resonant mode and to measure
cavity scattering parameters~\cite{SPARM1,SPARM2} at temperatures as
low as the critical temperature.
\begin{figure}[h]
  \centering
  \includegraphics[width=0.9\textwidth]{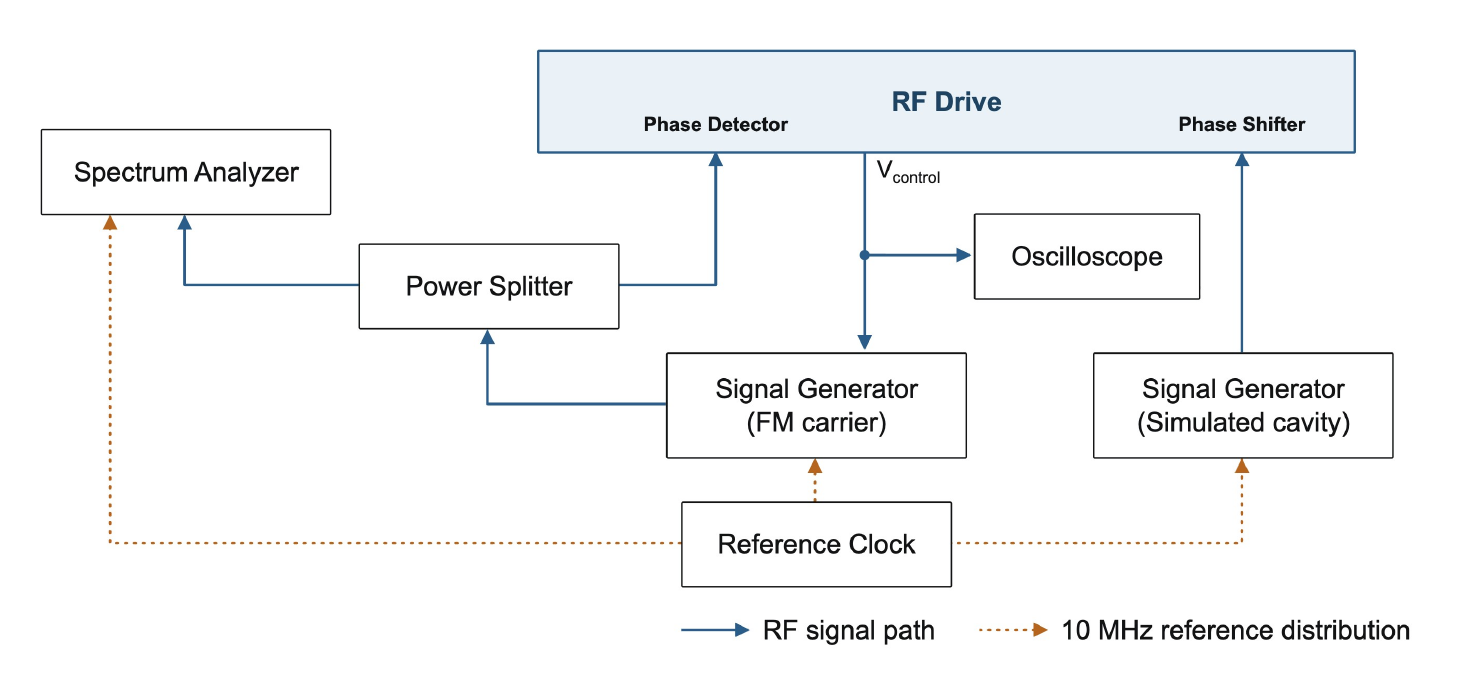}
  \caption{Test PLL configuration, with ``RF Drive'' corresponding to
    the portion indicated by green dashed lines in
    Figure~\ref{FIG:WHOLELLRF}.}
  \label{FIG:TESTPLL}
\end{figure}

\begin{figure}[h]
  \centering
  \includegraphics[width=0.85\textwidth]{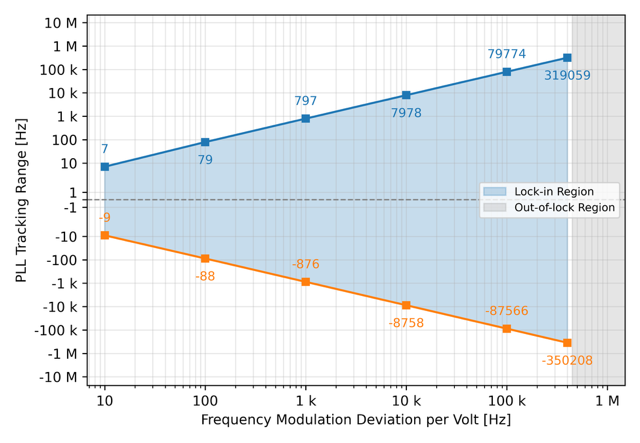}
  \caption{PLL tracking range as a function of FM deviation per unit
    voltage during testing.}
  \label{FIG:TESTRESULT}
\end{figure}

To demonstrate the locking behavior of the PLL developed to track the
SRF cavity's resonant frequency, a stand-alone LLRF test was conducted
using the experimental setup depicted in Figure~\ref{FIG:TESTPLL},
with the cryogenic system not yet implemented. In this configuration,
two signal generators were employed. One signal generator simulated
the SRF cavity's resonant frequency, whereas the other served as the
RF source, feeding its signal directly into the RF drive and operating
in frequency modulation (FM) mode so that its output frequency varied
in response to an external control voltage provided by the RF
drive. The RF signals from the generator operating in FM mode and the
generator simulating the cavity output were applied to the RF drive's
phase detector, which generated a voltage proportional to their phase
difference.
This voltage was then used to adjust the signal generator's output
frequency, forming the PLL feedback loop. The oscilloscope in the PLL
test setup measured the voltage-offset circuit's output voltages. For
the signal generator operating in FM mode, the FM deviation setting
determines the frequency offset per unit of the applied external
control voltage. Increasing the FM deviation setting generally widens
the achievable PLL frequency-tracking range. However, an excessive FM
deviation setting increases the loop gain, compromising loop
stability. Under such conditions, spurious tones appear, and the
spectral noise increases due to oscillations. As the FM deviation
setting approached the 1 MHz limit of our signal generator
(R\&S$\raisebox{0.25ex}{\circledR}$SMB100B~\cite{RSH}), spurious
components and spectral noise became apparent around the carrier,
limiting the test to an FM deviation setting of approximately 450~kHz.
Figure~\ref{FIG:TESTRESULT} presents the experimental results as a
function of FM deviation per unit voltage, with settings up to
approximately 450~kHz, illustrating a maximum stable operating range
of slightly above 300~kHz. The PLL achieved a tracking range of
approximately 300~kHz, more than sufficient to accommodate a reported
frequency fluctuation of 23~Hz due to microphonics in a
conduction-cooling system~\cite{MICROPHONICS}. The asymmetric tracking
range in Figure 6 stemmed from the phase detector's
frequency-dependent maximum output voltage~\cite{AD8302} and the
voltage-offset design based on an assumed maximum phase-detector
output voltage of 1.8~V. At the operating frequency of 1.5~GHz tested
in this study,\footnote{The operating frequency range is approximately
1.3$\--$1.6~GHz, limited by the RF filter
ZX75BP-1450-S+~\cite{MINICIRCUIT}.} the phase detector's maximum
output voltage was measured as 1.7~V using the oscilloscope via the
voltage-offset circuit; consequently, the FM input voltage ranged from
approximately $-0.9$ to $+0.8$~V, which is directly reflected in the
asymmetric tracking range presented in Figure~\ref{FIG:TESTRESULT}.
\subsection{Remote in situ SRF cavity measurements}\label{sec:power}
Once the RF drive described in Section~\ref{sec:pll} locks onto the
cavity's resonant frequency, the loaded quality factor $Q_L$ can be
measured using both the oscilloscope and spectrum analyzer depicted in
Figure~\ref{FIG:WHOLELLRF}, with the latter operating in zero-span
mode. The coupling coefficients $\beta^*_i$ introduced in
Equation~(\ref{EQ:BETA-2PORT}) can be obtained by measuring the
magnitude of the reflection coefficient at the relevant cavity port
$i$, $|\Gamma_i|$. $|\Gamma_i|$ satisfies the relation
$|\Gamma_i|=\sqrt{P_{i,\rm ref}/P_{i,\rm inc}}$, where
$P_{i,\rm ref}$ and $P_{i,\rm inc}$ are the reflected and incident
powers, respectively, at cavity port $i$ at the resonant
frequency.\footnote{Throughout this study, all measurements were
performed at the resonant frequency unless otherwise indicated by the
superscript "off."}
The input-port reflection-coefficient magnitude, $|\Gamma_{\rm in}|$,
can then be measured using the signal generator and two RF switches in the
configuration depicted in Figure~\ref{FIG:WHOLELLRF}; the two RF
switches were added to enable the use of the previously mentioned
network analyzer and remote in situ $|\Gamma_{\rm out}|$ measurements
discussed later in this section, rather than $|\Gamma_{\rm in}|$
measurements. The powers generated by the signal generator (sg) at on-
and off-resonance frequencies, $P_{\rm sg}$ and
$P^{\rm off}_{\rm sg}$, respectively, are given by
\begin{eqnarray}
  P_{\rm sg}=\frac{P_{\rm pm}L_{\rm in}}{G|\Gamma_{\rm in}|^2}\nonumber \\
  P^{\rm off}_{\rm sg}=\frac{P^{\rm off}_{\rm pm}L_{\rm in}}{G}, 
  \label{EQ:P_SG}
\end{eqnarray}
where $G$ is the SSPA gain and $L_{\rm  in}$ is the line attenuation
along the path from the signal generator to the power meter (pm) via the cavity
input port (see Figure~\ref{FIG:WHOLELLRF} and part of Figure~\ref{FIG:LLRF_BOTH}).
Therefore, $L_{\rm  in}$ includes attenuation from the lines installed
inside the refrigerator, which has the highest temperature gradient
within the entire conduction-cooling system and can thus be the most
susceptible to temperature instability and nonuniformity under ambient
environmental conditions.
Here, $P_{\rm pm}$ represents the portion of $P_{\rm sg}$ transmitted
to the cavity input port, partially reflected at that port depending
on $|\Gamma_{\rm in}|$, and subsequently transmitted to and measured
by the power meter. In contrast, at off-resonance frequencies,
$P^{\rm off}_{\rm sg}$ is transmitted to the cavity input port and fully
reflected because $|\Gamma_{\rm in}|=1$; the reflected power is then
transmitted to and measured by the power meter.
Setting $P_{\rm sg}=P^{\rm off}_{\rm sg}$ at the signal generator and
using Equation~(\ref{EQ:P_SG}) yields
\begin{eqnarray}
  |\Gamma_{\rm in}|=\sqrt{\frac{P_{\rm pm}}{{P^{\rm off}_{\rm pm}}}}.
  \label{EQ:GAMMA}
\end{eqnarray}
This remote in situ $|\Gamma_{\rm in}|$ measurement has typically been
performed without the additional RF switches mentioned earlier in this
section. The cavity input-port coupling coefficient
$\beta^*_{\rm in}$, introduced in Equation~(\ref{EQ:BETA-2PORT}), can
be determined using
\begin{eqnarray}
  \beta^*_{\rm in}&=&\frac{1+|\Gamma_{\rm in}|}{1-|\Gamma_{\rm in}|}=\frac{1+\sqrt{P_{\rm pm}/P^{\rm off}_{\rm pm}}}{1-\sqrt{P_{\rm pm}/P^{\rm off}_{\rm pm}}}>1 {\rm~for~the~overcoupling} \nonumber \\
           &=&\frac{1-|\Gamma_{\rm in}|}{1+|\Gamma_{\rm in}|}=\frac{1-\sqrt{P_{\rm pm}/P^{\rm off}_{\rm pm}}}{1+\sqrt{P_{\rm pm}/P^{\rm off}_{\rm pm}}}<1 {\rm~for~the~undercoupling}.
  \label{EQ:BETAfromGAMMA}
\end{eqnarray}
Here, we assumed that the gain $G$ and line attenuation $L_{\rm in}$
were the same at on- and off-resonance frequencies, an assumption that
is likely valid because the two frequencies can be selected very close
to each other owing to the SRF cavity's very narrow line shape.

By analogy with the $|\Gamma_{\rm in}|$ measurement, we enabled remote
in situ $|\Gamma_{\rm out}|$ measurement by adding the aforementioned
low-power circulator immediately after the cavity output port, thereby
introducing a new LLRF design for the first time in this study. This
design now allows remote in situ $|\Gamma_{\rm out}|$ measurement
without disconnecting, connecting, or reconnecting the relevant
components. The cavity output-port coupling coefficient
$\beta^*_{\rm out}$ can then be obtained using
Equations~(\ref{EQ:P_SG}), (\ref{EQ:GAMMA}), and
(\ref{EQ:BETAfromGAMMA}), with the gain terms in
Equation~(\ref{EQ:P_SG}) omitted because the SSPA is absent from the
relevant signal path.
The resulting $\beta^*_{\rm out}$ value is crucial for correctly
estimating $\beta_{\rm out}$ and, subsequently, $Q_0$ in
conduction-cooling systems, where temperature stability and uniformity
are relatively sensitive to ambient environmental conditions. Given
this precise measurement of $Q_0$ and the correct estimate of
$\beta_{\rm out}$, $Q_{\rm out}$ which is required to determine
$E_{\rm acc}$ (see Equation~(\ref{EQ:ECC})) can be obtained from the
relation $Q_{\rm out}=Q_0/\beta_{\rm out}$ in
Equation~(\ref{EQ:BETA0}).

\subsection{Remote in situ calibrations}\label{sec:calib}
Building on this new LLRF design for fully remote in situ SRF cavity
measurements described in Section~\ref{sec:power}, we also developed
remote in situ calibrations by adding a variable attenuator to improve
the precision of accelerating-gradient measurements. The power
transferred from the cavity output port, $P_{\rm out}$, satisfies the
relation
\begin{equation}
  P_{\rm out}=P^{\rm uncal}_{\rm out}L_{\rm cold}L_{\rm tocirc}L_{\rm hot},
  \end{equation}
where $P^{\rm uncal}_{\rm out}$ is the uncalibrated $P_{\rm out}$
transferred to and measured at the power meter; $L_{\rm cold}$ is the
line attenuation inside the refrigerator; $L_{\rm tocirc}$ is the line
attenuation from immediately outside the refrigerator to the low-power
circulator used for remote in situ $|\Gamma_{\rm out}|$ measurements,
as detailed in Section~\ref{sec:power}; and $L_{\rm hot}$ is the line
attenuation from immediately after the low-power circulator to the
power meter. In conduction-cooling systems, the line section
corresponding to $L_{\rm cold}$ has the highest temperature gradient
in the entire system; its temperature stability and uniformity can
thus be the most susceptible to ambient environmental effects, unlike in
convection-cooling systems using costly LHe, as discussed in
Section~\ref{sec:power}. Therefore, in situ $P_{\rm out}$ measurements
in conduction-cooling systems are particularly necessary and desirable
because they effectively correspond to in situ $L_{\rm cold}$
calibrations.

To extract $P_{\rm out}$ from $P^{\rm uncal}_{\rm out}$, we also used
the signal generator, an RF switch, and an additional variable
attenuator in the arrangement depicted in
Figure~\ref{FIG:WHOLELLRF}. The power generated by the signal
generator at the resonant frequency is given by
\begin{equation}
P_{\rm sg}=\frac{P_{\rm pm}L_{\rm var}L^2_{\rm cold}L^2_{\rm tocirc}L_{\rm hot}}{|\Gamma_{\rm out}|^2},
\end{equation}
where $P_{\rm pm}$ is the portion of $P_{\rm sg}$ transferred to the
cavity output port, partially reflected at that port depending on
$|\Gamma_{\rm out}|$, and subsequently transferred to and measured at
the power meter. $L_{\rm var}$ is the variable line attenuation from
the signal generator to the low-power circulator, adjusted using the
variable attenuator. By adjusting $L_{\rm var}$ to match
$L_{\rm hot}$, we can obtain $P_{\rm out}$ from the relation
\begin{equation}
  P_{\rm out}=P^{\rm uncal}_{\rm out}|\Gamma_{\rm out}|\sqrt{\frac{P_{\rm sg}}{P_{\rm pm}}},
\end{equation}
where we again assumed equal line attenuation at on- and off-resonance
frequencies, as in the $|\Gamma_{\rm out}|$ measurements described in
Section~\ref{sec:power}. This procedure enables remote in situ
$P_{\rm out}$ calibration from outside the high-power region to avoid
exposure to hazardous radiation.

\section{Control and DAQ software}\label{sec:daq}
Operating an LLRF system for conduction-cooled SRF cavity measurements
imposes two control-software requirements uncommon in a laboratory
benchtop setup. First, because the RF drive, circulator introduced in
this study, and diagnostic instruments (spectrum analyzer, network
analyzer, oscilloscope, and power meters) must be used together in a
reproducible sequence to perform the PLL-locking and in situ
measurement and calibration procedures described in
Section~\ref{sec:llrf}, the software must provide access to the
relevant heterogeneous, commercially sourced instruments through a
single coherent interface rather than a collection of vendor-specific
tools.
Second, because conduction-cooling systems are more exposed to ambient
temperature fluctuations than LHe-bath systems and because the
high-power region of the setup is a radiation area during cavity
operation, the system must support operation and continuous monitoring
from outside the experimental hall. These requirements motivated the
development of dedicated \verb|Python|-based control and DAQ software
named \verb|kullrf| alongside the LLRF hardware.
\begin{figure}[h]
  \centering
  \includegraphics[width=0.85\textwidth]{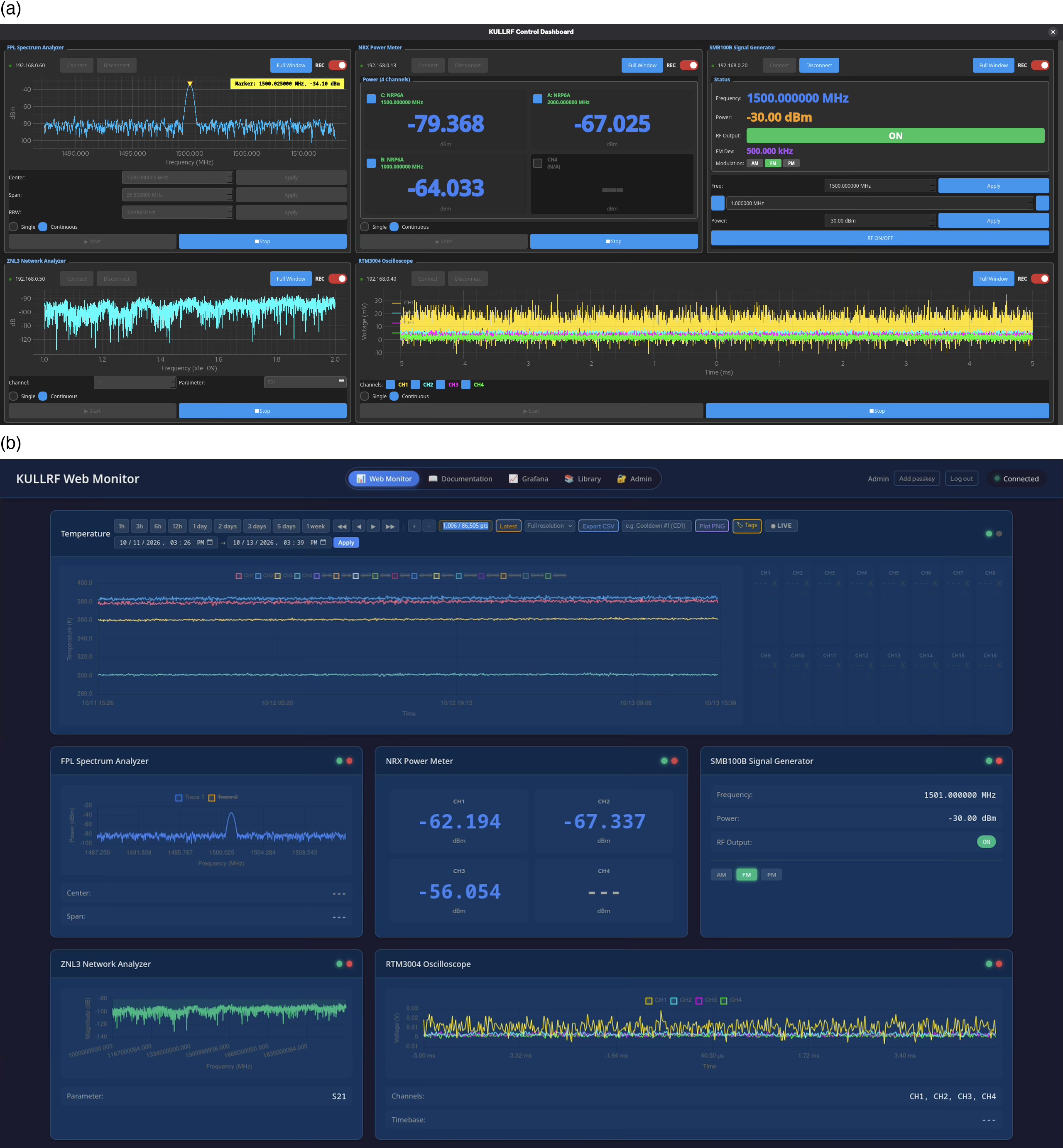}
  \caption{Software interfaces of the LLRF system. (a) PyQt6-based
    graphical user interface (GUI) for instrument control and
    monitoring; (b) FastAPI-based web interface for remote
    monitoring.}  
  \label{FIG:DAQ}
\end{figure}
\verb|kullrf| comprises three layers: core, device, and
application; each successive layer builds on the one below it. The
core layer provides infrastructure shared by every instrument:
connection handling through the Virtual Instrument Software
Architecture (VISA) using the \verb|pyvisa| library \cite{pyvisa} over
Ethernet, with RS-232 and the General Purpose Interface Bus retained
for backward compatibility; Modbus communication over the Transmission
Control Protocol for the temperature sensors installed on the system;
and a local \verb|SQLite|~\cite{sqlite} database that persistently
records every measurement. Built on the core layer, the device layer
represents each instrument in Figure~\ref{FIG:WHOLELLRF} (the spectrum
analyzer, network analyzer, oscilloscope, power meters, and signal
generator) through a dedicated \verb|Python| module mapping the
manufacturer's native Standard Commands for Programmable Instruments
command set to validated, intuitively named functions. This separation
isolates the rest of the software from instrument- and
communication-protocol-specific details and allows a new instrument to
be added without changing how the application layer operates it; the
same modular approach has already been reused across 10 instrument
types in this and related setups at our institute.

The application layer, in turn, provides access to the full device
layer through two complementary front ends. A \verb|PyQt6|-based desktop
application~\cite{pyqt6} (see Figure~\ref{FIG:DAQ} (a)) provides an
operator with direct, real-time control of every instrument for setup,
tuning, and calibration, whereas a separate FastAPI-based web
monitor~\cite{fastapi} (see Figure~\ref{FIG:DAQ} (b)), accessible over
the network with per-user authentication, streams live measurements to
any browser via \verb|WebSocket| without requiring access to the
personal computer used for control. The two front ends do not
communicate directly; instead, every measurement is recorded in the
core layer's database, from which the web monitor retrieves the
data. This decoupling keeps the remote monitor operational and the
measurement history intact even if the desktop application is
restarted, the operator is away, or the control PC is power-cycled;
this capability is important in a radiation-controlled area where the
instrument rack cannot be freely accessed while RF power is applied to
the cavity. The same database also supports long-term recording to
track temperature stability and uniformity across the
conduction-cooling chain.
As discussed in Section~\ref{sec:intro}, limitations in these
properties constitute the central environmental concern that in situ
measurements and calibrations are designed to compensate for. These
histories can be inspected directly or exported to
\verb|Grafana|~\cite{grafana} for extended trend analysis and
cross-checking against the calibration introduced in Section
2.3~\ref{sec:calib}.

Although \verb|kullrf| was developed specifically for this LLRF
system, its device and application layers are not limited to any
single measurement procedure. We expect the same software to be
extended to support DAQ for the SRF cavity measurement itself and,
more broadly, to support other conduction-cooled or remotely operated
RF systems at institutes requiring coordinated multi-instrument
control and continuous, unattended remote monitoring.

\section{Summary}\label{sec:daq}
We developed an LLRF system for SRF cavity measurements in a
conduction-cooling system. We adopted a VCO as part of a PLL to track
the frequencies of SRF cavity modes. The system uses
{\tt{Python}}-based software developed in-house. We achieved a maximum
stable PLL operating range of slightly above 300 kHz, more than
sufficient to accommodate typical frequency fluctuations due to
microphonics in conduction-cooling systems.

By incorporating the additional circulator described above, we
developed a new LLRF system design enabling fully remote in situ
measurements and calibrations for the first time, primarily to ensure
safety but also to improve measurement precision; improved precision
is crucial in conduction-cooling systems, where temperature stability
and uniformity are poorer than those in convection-cooling systems
using costly LHe.

Although our LLRF system was developed for SRF cavity measurements in
a conduction-cooling system, it can be used for any other SRF cavity
measurement.

\section*{Declaration of Generative AI Use}
Part of the \verb|kullrf| control and DAQ software described in this
work was developed with the assistance of Claude Code (Anthropic), an
AI coding assistant.

\section*{Acknowledgments}
This work was supported by a Korea University Grant, the National
Research Foundation of Korea (NRF) grants funded by the Korea
government (MSIT) (RS-2025-00556247, RS-2022-00143178, and
RS-2025-23963275), and the Korea Basic Science Institute (National
research Facilities and Equipment Center) grant funded by the Korea
government (MSIT) (NFEC-2019R1A6C1010027).

\end{document}